\documentstyle[preprint,prc,aps,floats,epsf]{revtex}

\tightenlines

\begin{document}

% read in symbols
%\input eft_symbols.tex
%
% Begin: Simple substitution macros used in the text
%
\def\symdef#1#2{\def#1{#2}}
%
%\begin{symbols}

\symdef{\alphabar}{\overline\alpha}
\symdef{\alphabarp}{\overline\alpha\,{}'}
\symdef{\Azero}{A_0}

\symdef{\bzero}{b_0}

\symdef{\Cs}{C_{\rm s}}
\symdef{\cs}{c_{\rm s}}
\symdef{\Cv}{C_{\rm v}}
\symdef{\cv}{c_{\rm v}}

\symdef{\dalem}{\frame{\phantom{\rule{8pt}{8pt}}}}
\symdef{\del}{\partial}
\symdef{\delN}{{\wt\partial}}
\symdef{\Deltaevac}{\Delta{\cal E}_{\rm vac}}

\symdef{\ed}{{\cal E}}
\symdef{\edens}{{\cal E}}
\symdef{\edk}{{\cal E}_k}
\symdef{\edkzero}{{\cal E}_{k0}}
\symdef{\edv}{{\cal E}_{\rm v}}
\symdef{\edvphi}{{\cal E}_{{\rm v}\Phi}}
\symdef{\edvphizero}{{\cal E}_{{\rm v}\Phi 0}}
\symdef{\edzero}{{\cal E}_{0}}
\symdef{\Efermistar}{E_{{\scriptscriptstyle \rm F}}^\ast}
\symdef{\Efermistarzero}{E_{{\scriptscriptstyle \rm F}0}^\ast}
\symdef{\etabar}{\overline\eta}
\symdef{\ezero}{e_0}

\symdef{\fomega}{f_\omegav}
\symdef{\fpi}{f_\pi}
\symdef{\fv}{f_{\rm v}}
\symdef{\fvt}{\widetilde\fv}

\symdef{\gA}{g_A}
\symdef{\gammazero}{\gamma_0}
\symdef{\gomega}{g_\omegav}
\symdef{\gpi}{g_\pi}
\symdef{\grho}{g_\rho}
\symdef{\grad}{{\bbox{\nabla}}}
\symdef{\gs}{g_{\rm s}}
\symdef{\gssq}{\gs^2}
\symdef{\gv}{g_{\rm v}}
\symdef{\gvsq}{\gv^2}

\symdef{\fm}{\mbox{\,fm}}

\symdef{\infm}{\mbox{\,fm$^{-1}$}}
\symdef{\isovectorTensor}{s_{\tauvec}}
\symdef{\isovectorTensorN}{\wt\isovectorTensor}
\symdef{\isovectorVector}{j_{\tauvec}}
\symdef{\isovectorVectorN}{\wt j_{\tauvec}}

\symdef{\kappabar}{\overline\kappa}
\symdef{\kfermi}{k_{{\scriptscriptstyle \rm F}}}
\symdef{\kfermizero}{k_{{\scriptscriptstyle \rm F}0}}
\symdef{\Kzero}{K_0}

\symdef{\lambdabar}{\overline\lambda}
\symdef{\LdotS}{\bbox{\sigma\cdot L}}
\symdef{\lsim}{\lower0.6ex\vbox{\hbox{$\ \buildrel{\textstyle <}
         \over{\sim}\ $}}}
\symdef{\lzero}{l_{0}}

\symdef{\Mbar}{\overline M}
\symdef{\Mbarzero}{\Mbar_0}
\symdef{\MeV}{\mbox{\,MeV}}
\symdef{\momega}{m_\omegav}
\symdef{\mpi}{m_\pi}
\symdef{\mrho}{m_\rho}
\symdef{\ms}{m_{\rm s}}
\symdef{\mssq}{\ms^2}
\symdef{\Mstar}{M^\ast}
\symdef{\Mstarzero}{M^\ast_0}
\symdef{\mv}{m_{\rm v}}
\symdef{\mvsq}{\mv^2}
\symdef{\mzero}{{\rm v}_{0}}

\symdef{\Nbar}{$\overline{\rm N}$\ }
\symdef{\NN}{NN\ }     % roman
%\symdef{\NN}{$NN$\ }   % italics
\symdef\NNbar{$\overline{\rm N}$N\ }

\symdef{\omegaV}{V}
\symdef{\omegav}{{\rm v}}

\symdef{\Phizero}{\Phi_0}
\symdef{\psibar}{\overline\psi}
\symdef{\psidagger}{\psi^\dagger}
\symdef{\pvec}{{\bf p}}

\symdef{\rhoB}{\rho_{{\scriptscriptstyle \rm B}}}
\symdef{\rhoBt}{\wt\rho_{{\scriptscriptstyle \rm B}}}
\symdef{\rhoBzero}{\rho_{{\scriptscriptstyle \rm B}0}}
\symdef{\rhominus}{\rho_{-}}
\symdef{\rhoplus}{\rho_{+}}
\symdef{\rhos}{\rho_{{\scriptstyle \rm s}}}
\symdef{\rhospzero}{\rho'_{{\scriptstyle {\rm s} 0}}}
\symdef{\rhost}{\wt\rho_{{\scriptstyle \rm s}}}
\symdef{\rhoszero}{\rho_{{\scriptstyle {\rm s}0}}}
\symdef{\rhotau}{\rho_{\tauvec}}
\symdef{\rhotaut}{\wt\rho_{\tauvec}}
\symdef{\rhothree}{\rho_{3}}
\symdef{\rhothreet}{\wt\rho_{3}}
\symdef{\rhozero}{\rho_0}

\symdef{\scalar}{\rhos}
\symdef{\scalarN}{{\rhost}}
\symdef{\Szero}{S_0}

\symdef{\tauvec}{{\bbox{\tau}}}
\symdef{\tauthree}{\tau_3}
\symdef{\tensor}{{{s}}}
\symdef{\tensorN}{\wt\tensor}
\symdef{\tensort}{\wt{{s}}}
\symdef{\tensorthree}{\tensor_3}
\symdef{\tensorthreet}{\wt\tensorthree}
\symdef{\Tr}{{\rm Tr\,}}

\symdef{\umu}{u^\mu}
\symdef{\Ualpha}{U_{\alpha}}
\symdef{\Ueff}{U_{\rm eff}}
\symdef{\Uzero}{U_0}
\symdef{\Uzerop}{U_0'}
\symdef{\Uzeropp}{U_0''}

\symdef{\vecalpha}{{\bbox{\alpha}}}
\symdef{\veccdot}{{\bbox{\cdot}}}
\symdef{\vecnabla}{{\bbox{\nabla}}}
\symdef{\vecpi}{{\bbox{\pi}}}
\symdef{\vectau}{{\bbox{\tau}}}
\symdef{\vector}{j_{\scriptscriptstyle V}}
\symdef{\vectorN}{{\wt\vector}}
\symdef{\vecx}{{\bf x}}
\symdef{\Vopt}{V_{\rm opt}}
\symdef{\Vzero}{V_0}

\symdef{\wt}{\widetilde}
\symdef{\wzero}{w_0}
\symdef{\Wzero}{W_0}

\symdef{\zetabar}{\overline\zeta}
%
%
%
% End: Simple substitution macros
%
%\end{symbols}

% other definitions
\def\beq{\begin{equation}}
\def\eeq{\end{equation}}
\def\beqa{\begin{eqnarray}}
\def\eeqa{\end{eqnarray}}

% uncomment \draft to have PACS numbers appear
\draft

% put preprint numbers.  
\preprint{IU/NTC\ \ 00--03}

\title{Quantum Hadrodynamics: Evolution and Revolution}

\author{R. J. Furnstahl} 

\address{Department of Physics \\
         The Ohio State University,\ \ Columbus, OH\ \ 43210}
\author{Brian D. Serot}
\address{Department of Physics and Nuclear Theory Center \\
         Indiana University,\ \ Bloomington, IN\ \ 47405}
\date{June, 2000}
\maketitle
\begin{abstract}
The underlying philosophy and motivation for quantum hadrodynamics (QHD),
namely, 
relativistic field theories of nuclear phenomena featuring manifest covariance, 
have evolved over the
last quarter century in response to successes, failures, and sharp criticisms.
A recent revolution in QHD,
based on modern
effective field theory and density functional theory perspectives,
explains the successes, provides antidotes to the failures,
rebuts the criticisms, and focuses
the arguments in favor of a covariant representation. 
%\end{abstract}
%

\end{abstract}

\bigskip
\pacs{Submitted to {\it Comments on Modern Physics}}

%%%%%%%%%%%%%%%%%%%%%%%%%%%%%%%%%%%%%%%%%%%%%%%%%%%%%%%%%%%%%%%%%%%%%%
% Switch to symbols for footnotes
 \renewcommand{\thefootnote}{\fnsymbol{footnote}}
 \addtocounter{footnote}{1}    % start with dagger, not asterisk
%%%%%%%%%%%%%%%%%%%%%%%%%%%%%%%%%%%%%%%%%%%%%%%%%%%%%%%%%%%%%%%%%%%%%%

\section{Introduction}

{\it Quantum hadrodynamics\/} (QHD) refers to relativistic field theories 
for nuclei based on hadrons,  in which the representation
is manifestly covariant.
The distinguishing empirical feature of covariant QHD
is the presence of large (several hundred MeV), isoscalar, 
Lorentz scalar and vector mean fields (optical potentials) in nuclear
matter at normal nuclear densities.
During the last quarter century, QHD calculations have had 
numerous successes, but  
also apparent failures 
that arose when the dynamics of the quantum vacuum were
computed with the same degrees of freedom used to
describe the valence-nucleon physics.
Moreover, there have been sharp criticisms over the years (some of
which appeared in earlier issues of {\it Comments\/}) based on the
relevance of a covariant approach to the nuclear many-body problem at
observable densities, the use of relativistic, renormalizable
quantum field theories with nucleon fields 
to describe composite hadrons and the quantum chromodynamics 
(QCD) vacuum,
and the apparent lack of pion dynamics and chiral symmetry in the 
empirically successful calculations.

The underlying philosophy of QHD has evolved over the past twenty-five years
in response to these successes, failures, and criticisms.
Ultimately, the original motivation for 
renormalizable QHD lagrangians yielded to a more general approach
based on the modern ideas of nonrenormalizable, effective 
relativistic field theories.
This change in QHD philosophy  is nothing short of
a ``revolution'' that occurred during a two- or three-year span in 
the mid 1990s.
The new approach, based on effective field theory
(EFT) and density functional theory (DFT), allows us to 
understand the successful
mean-field calculations of nuclear properties
and how chiral symmetry works in QHD.
It provides antidotes to earlier failures through a 
consistent, systematic, covariant treatment of the nuclear many-body
problem, and it provides rebuttals to the previous criticisms.

In this {\it Comment}, we trace the QHD evolution and revolution,
with special attention to how past criticisms and deficiencies have
been nullified.

\section{Evolution}

\subsection{Original Motivation}
 
Calculations of nuclear many-body systems based on relativistic
hadronic field theories have existed for many years; one can trace
their history back at least as far as the seminal work of Schiff in 
1951 \cite{SCHIFF51}.
An important advance was made roughly 25 years ago, when Walecka
enumerated the philosophy underlying a covariant description of
nuclear matter, together with the rules for calculating within such
a framework and a systematic program for investigating the observable
consequences of the approach \cite{WALECKA74}.%
\footnote{At the time of Walecka's original paper, several other groups 
were
studying relativistic, hadronic, field-theoretic approaches to the
nuclear many-body problem, using similar but slightly different
philosophies.
See Ref.~\protect\cite{SEROT86} for discussions of other approaches 
and results.}

To paraphrase the motivation presented so clearly by Walecka: 
to discuss neutron stars, it is necessary to have an equation
of state that describes matter from observed terrestrial densities
upward; a consistent theory should include mesonic degrees of freedom
explicitly to allow for extrapolation to high densities; as the 
density
of the matter is increased, relativistic propagation of the nucleons
(and the retarded propagation of the virtual mesons) must be included;
and causal restrictions on the propagation of excitation modes of the
interacting system must be automatically contained in the theory.
The basic conclusion following from these ideas is that, ``The only
consistent approach \dots\ which meets these [objectives] is \dots\ 
a local, relativistic, many-body quantum field theory.''

The systematic program to develop QHD as defined by Walecka focused
on the nuclear many-body problem and relied on {\it renormalizable\/}
lagrangian densities to define the models.
This restriction was motivated by the desire
to extrapolate away from
the empirical calibration data in a manner that did not introduce
any new, unknown parameters.
Perhaps more importantly, the faithful pursuit of this framework would
reveal whether renormalizable QHD was feasible and practical or 
not \cite{SEROT86,SEROT91}.

The original model contained neutrons, protons, and isoscalar,
Lorentz scalar and vector mesons.
In the nonrelativistic (Yukawa potential) limit (which was never
actually used in the calculations), single-meson exchange
generates basic observed features of the \NN interaction: a
strong, short-range repulsion and a medium-range attraction.
It was assumed by fiat that the possible (renormalized) nonlinear 
couplings between the scalar fields were zero, for simplicity, 
although this restriction was relaxed quite early in the development
by other practitioners.
Moreover, no attempt was made to reconcile the model with the 
spontaneously broken, approximate 
$SU(2)_L \times SU(2)_R$ chiral symmetry of hadronic interactions.

The model contains large couplings, so a 
practical, nonperturbative approximation is needed as a starting
point to describe the
nuclear equation of state (EOS).
Walecka argued that at high enough density, fluctuations in the 
meson fields could be ignored, and they could be replaced by their
classical expectation values or {\it mean fields}.
He also assumed that these conditions were 
sufficiently valid at ordinary nuclear
density, so that the model could be calibrated in the mean-field
approximation, and that the mean-field contributions would dominate
the high density (e.g., neutron-star) EOS.
Corrections to the mean-field approximation were calculated, and 
it was indeed
found that in the context of renormalizable, Walecka-type models,
the ``stiff'' EOS predicted by the mean-field theory (MFT),
in which the pressure approaches the total energy density 
from below, becomes increasingly
accurate as the nuclear density increases.

The hope was that nonrenormalizable and vacuum (short-range) effects,
which had not been included in the corrections noted above,
would be small enough to be described adequately by
the long-range degrees of freedom 
of a renormalizable field theory, through
the systematic evaluation of quantum loops.
This hope was not fulfilled
by explicit calculation of short-range loop effects.
Furthermore,
enlarging the nonlinear meson self-couplings to allow for effective, 
nonrenormalizable terms alters the high-density
nuclear EOS qualitatively \cite{MUELLER96}.

\subsection{Successes}

At the same time as the neutron-matter EOS was being studied, the
MFT model was applied to the bulk and single-particle properties
of doubly magic nuclei.
The most important conclusion of this early work was that calibration
to the empirical equilibrium point of ordinary nuclear matter
produces scalar and vector mean fields of roughly several hundred
MeV at equilibrium density. 
When extended to finite nuclear systems, the resulting
single-particle spin-orbit potential is roughly the same size as
the observed spin-orbit potential, and thus the relativistic MFT 
{\it predicts the existence of the nuclear shell model}, 
without any {\it ad hoc\/} adjustments to the spin-orbit force.
Moreover, MFT models that incorporated nonlinear (i.e., cubic and
quartic) scalar field couplings reproduced bulk and single-particle
nuclear observables as well as or better than any other concurrent
models.

Relativistic MFT calculations have been
performed for nuclei throughout the Periodic Table, with similarly
realistic results and predictions; the reader is directed to the
cited review articles for discussions of these numerous calculations
\cite{HOROWITZ81,REINHARD86,RUFA88,REINHARD89,GAMBHIR90,SEROT92,SEROT97}.
When the MFT nuclear densities were folded with the free \NN scattering
matrix to compute proton--nucleus scattering observables (this is
called the relativistic impulse approximation or RIA
\cite{MCNEIL83,SHEPARD83,CLARK83a,CLARK83b,CLARK84a}), excellent
descriptions of existing data and predictions for upcoming data were
found---far superior to nonrelativistic calculations at the same
level of approximation.

The vast majority of successful QHD predictions rely on the
important observation that there are large, isoscalar, Lorentz
scalar and vector mean fields in nuclei.
Moreover, the successful calculations explicitly
include only valence nucleons
and long-range, many-body dynamics; the QHD degrees of freedom are
designed precisely to describe this type of dynamics.
It is also possible to study nuclear excited states in a
random-phase approximation (RPA) that involves only long-range
dynamics, but that still maintains the underlying symmetries of the 
lagrangian \cite{DAWSON90}.

To explicitly include two-nucleon correlations, one uses the so-called
Dirac--Brueckner--Hartree--Fock (DBHF) theory.
With a covariant \NN kernel that is fitted
to two-body data and that contains
large Lorentz scalar and vector (and pionic) components, one can 
simultaneously reproduce the nuclear matter equilibrium point at the 
two-hole-line level \cite{BROCKMANN84,MACHLEIDT89,DEJONG91}.
Moreover, although the correlation corrections produce changes in the MFT
binding energy that are of the same order as the binding energy itself,
{\it the corrections to the large MFT scalar and vector self-energies
(optical potentials) are small\/} \cite{HOROWITZ84,HOROWITZ87}.
Nevertheless, to our knowledge, numerous approximations in the DBHF approach
have never been quantitatively tested, and the systematic inclusion of
contributions from the quantum vacuum was an unsolved problem
\cite{SEROT92}.

\subsection{Difficulties}

Along with the numerous successes, there were also various
difficulties that can be traced to a
common source: the requirement that the QHD models be 
{\it renormalizable}.
The difficulties fall into two basic classes: those arising from the
computation of quantum loops (short-range physics) using long-range
degrees of freedom, and those arising from attempting to maintain both
renormalizability and chiral symmetry simultaneously.

Although the relativistic MFT results are encouraging, the QHD
program sought a more complete
description of the nuclear many-body system, which {\em requires\/} 
the development of
reliable techniques to extend these calculations.
Quantum loops are important for several reasons: loops ensure the 
unitarity of scattering amplitudes, baryon loops containing valence
nucleons
incorporate familiar many-body effects, loops introduce effects 
arising from the modification of the quantum vacuum in the presence of
valence nucleons, and meson loops in particular generate contributions
to the extended structure of the nucleon.

Not all loops in QHD are problematic.
For example, loops involving fluctuations of the pion field generate
long-range effects (since the pion is light) and should
be accurately described by explicit calculation within the QHD model
(provided that approximate chiral symmetry is maintained).
The strong,
mid-range \NN attraction arises from pion rescattering loops when
two pions in the scalar, isoscalar channel are exchanged between 
nucleons.
These long-range dynamical effects are much more efficiently described
with hadrons than with QCD quarks and gluons; in fact, most QHD models
go one step further and simulate the mid-range \NN attraction using 
a Yukawa coupling to an explicit scalar, isoscalar field.

Problems arise with loops when one attempts to describe short-range
dynamics using the heavier QHD degrees of freedom (nucleons and
non-Goldstone bosons).
In a renormalizable theory, a finite result can be obtained for the
Casimir effect, and its addition to the MFT produces what is usually
called the relativistic Hartree approximation (RHA).
Although the new contributions are finite,  
they degrade the agreement of the nuclear
predictions with experiment, particularly when one examines spin-orbit
splittings for single-particle levels near the Fermi surface.
These results imply that the QHD treatment of the
quantum vacuum at the one-baryon-loop level is, at best, inadequate;
although higher-order corrections might reduce the size of the one-loop
terms, this can occur only through sensitive cancellations between
relatively large contributions.

In fact, contributions from higher-loop terms within the 
renormalizable
QHD framework do not improve the situation.
Explicit calculations of the nuclear matter energy density at the
two-loop level found enormous contributions that altered the 
description of the nuclear ground state 
qualitatively \cite{TWOLOOP89}.
The conclusion was that the loop expansion does not provide a reliable
approximation scheme in renormalizable QHD.

Similarly, vacuum contributions in the summation of ring diagrams
produced unphysical poles at spacelike momenta in the meson 
propagators
(sometimes called ``ghosts''), which signal either an inconsistency
in the QHD framework or an inadequate level of approximation.
It was proposed that vertex corrections within the theory could solve
these problems \cite{MILANA91,ALLENDES92}, 
but complete calculations involving vertex insertions proved
to be impractical, and to our knowledge, no systematic, reliable
approximation scheme for incorporating both long-range and short-range
loop effects in renormalizable QHD theories has ever been found.

The second class of difficulties arises when one tries to embed the
approximate, spontaneously broken, $SU(2)_L \times SU(2)_R$
chiral symmetry of QCD in a renormalizable QHD theory.
The original Walecka model made no mention of chiral symmetry, but
there were chiral models in use at that time based on the well-known
Sigma model of Schwinger \cite{SCHWINGER57}
and of Gell-Mann and L{\'e}vy \cite{SIGMA60}.
In fact, if one takes the original Walecka model, adds massless
pions that couple to nucleons with a pseudoscalar ($\gamma_5$)
Yukawa coupling, and demands that the theory be Lorentz covariant,
parity invariant, isospin and chiral invariant, and renormalizable,
one is led to a lagrangian that is simply the Sigma model
with an additional isoscalar vector meson.
The nonzero nucleon and scalar masses are generated through the
familiar spontaneous symmetry breaking, and given the similarity to
the Walecka-model lagrangian, it is natural 
to  identify the scalar field (which is the chiral partner of the
pion) with the scalar field in the Walecka model.

The MFT for the chiral model can be motivated precisely as before,
except that there are now cubic and quartic scalar self-couplings that
are not free parameters, but that are specified by the chiral 
symmetry.
Unfortunately, the assumption that the chiral scalar field is the same
as the Walecka scalar leads to dire consequences.
First, it is impossible to reproduce the empirical nuclear matter
equilibrium point in the MFT.
Including quantum loops does not help, since one either generates
extremely large contributions or
arrives at uncertain results due to the appearance of unphysical
poles in the meson propagators.
An extensive mean-field analysis shows that it is 
impossible to generate realistic results for finite nuclei within
this framework (see Refs.~\cite{FURNSTAHL93a,FURNSTAHL93b,FURNSTAHL96}).
The conclusion is that the standard form of spontaneous chiral 
symmetry breaking, implemented in a model with a linear realization
of the symmetry, {\it cannot\/} produce successful nuclear
phenomenology at the mean-field level, if the chiral scalar field is
identified with the scalar field in the Walecka model.
This failure of the Sigma model is evidence that the simultaneous
constraints of renormalizability and linear chiral symmetry are
too restrictive.

Some progress was made on this problem by following the work of
Weinberg and by making field transformations of the nucleon, pion, 
and scalar fields \cite{WEINBERG67}.
In addition to changing the form of the pion--nucleon interaction to
include a pseudovector ($\gamma_{\mu} \gamma_5$) coupling, 
the transformation allows the introduction of a {\it new\/}
scalar, isoscalar field that is not the chiral partner of the pion.
This field  plays the same role as in the
Walecka model: it simulates important $\pi\pi$ and \NN interactions
that must be included from the outset to generate a realistic 
description of nuclear matter and nuclei.

A profound change has occurred, however, because in contrast to the
original proposal of QHD as a renormalizable field theory, we are now
forced to consider the new scalar field as an effective degree of
freedom and the new chiral lagrangian as a nonrenormalizable, 
effective lagrangian.
Thus the earlier philosophy of QHD must be generalized
to include nonrenormalizable, effective field theories.

\section{Revolution}

\subsection{EFT/DFT Perspective}

The revolution in QHD started with the reinterpretation of QHD lagrangians
as nonrenormalizable EFT lagrangians.
An effective lagrangian consists of known long-range interactions constrained
by symmetries and a complete set of generic short-range interactions.
The division between  long and short
is characterized by the breakdown scale $\Lambda$ of the EFT.
While it is not possible at present to derive an effective
hadronic theory directly from the underlying QCD, 
the EFT perspective implies that this is not necessary.
If one constructs 
a general lagrangian that respects the symmetries of QCD: Lorentz
covariance, parity conservation, time-reversal and charge-conjugation
invariance, (approximate) isospin symmetry, and 
spontaneously broken chiral symmetry,
then the EFT is a general parametrization of observables
below the breakdown scale.

The EFT perspective, with the freedom to redefine and transform fields,
implies that {\it there are infinitely many  representations
of  low-energy QCD physics\/}.
But they are not all equally efficient or physically transparent.
One of the possible choices is between Lorentz covariant
and nonrelativistic formulations.
(In the context of EFT, these can be related by the heavy-baryon
expansion \cite{JENKINS91}.)
Recent developments in baryon chiral perturbation theory support the
consistency (and utility) of a covariant EFT,
with Dirac nucleon fields in a Lorentz invariant effective 
lagrangian density \cite{ELLIS98,BECHER99}.
A similar framework underlies QHD approaches to nuclei.

For QHD, we identify $\Lambda$ with the scale of non-Goldstone-boson
physics (roughly 600\MeV).
At momenta small compared to $\Lambda$, short-distance physics (such as
the substructure of nucleons) is only partially resolved
and so may be incorporated into the coefficients of operators
organized as a derivative expansion.
The coefficients of these short-range terms may eventually be derived from
QCD, but at present, they must be fitted by matching calculated and
experimental observables.
In principle, there are an infinite number of 
(nonrenormalizable) terms, but in practice, the
lagrangian or energy functional can be truncated to work to a 
given precision \cite{FURNSTAHL97}.
The EFT is useful if this truncation can be made at low enough order that
the number of free parameters is not prohibitive.

In QHD, the only {\it essential\/} hadronic degrees of freedom are 
the nucleons and pions.
The long-range
pion--pion and pion--nucleon interactions are included in a nonlinear
realization of chiral symmetry, which avoids dynamical assumptions 
inherent in linear representations.
These interactions can be written down systematically (given a 
power-counting scheme) \cite{FURNSTAHL97}.
Low-mass vector mesons are typically included for phenomenological reasons,
but are not required since their masses are of the order of $\Lambda$;
they are absent from point-coupling models, for example.
In descriptions of \NN scattering and of nuclear structure and 
reactions, the heavy bosons carry {\it spacelike\/}
four-momenta and are ``off the mass shell''; they therefore serve
simply as a convenient way to parametrize the \NN interaction in
exchange channels with vector quantum numbers.
This explains why it is useful to introduce collective degrees of freedom
with other quantum numbers, such as a 
Delta baryon (with spin and isospin of 3/2) to incorporate 
important pion--nucleon interactions. 
Because one must always truncate the lagrangian,
these degrees of freedom can be efficient in the many-body problem
{\em whether or not they are actually observed as hadronic 
resonances\/}.

A scalar, isoscalar
mean-field in nuclei is an efficient way to include implicitly the
effects of pion exchange that are the most important for describing
bulk nuclear properties.
Because chiral symmetry is realized nonlinearly, one
can {\it add\/} a light scalar, isoscalar,
chiral-singlet field to the theory and give
it a Yukawa coupling to the nucleon, just as in the Walecka model.
Nonlinear self-interactions of this new scalar must be included,
with adjustable couplings that arise in part from the nucleon substructure.
Since the expectation value of the pion field in nuclear matter 
vanishes
at the mean-field level, one makes the remarkable observation that the
MFT of Walecka-type QHD models is the same as the
MFT of the  chiral EFT model!
Thus the MFT of Walecka-type models {\it is consistent with chiral
symmetry}, provided we think in terms of a nonlinear realization of 
the symmetry.
The light scalar, isoscalar field, which is {\it not\/}
the chiral partner of the pion, plays the same role as in the
Walecka model: it simulates important $\pi\pi$ and \NN interactions
that must be included from the outset to generate a realistic 
description of nuclear matter and nuclei.

To make systematic calculations,
the EFT approach exploits the separation of scales
in physical systems, with 
the ratios of scales providing expansion parameters.
A connection between appropriate QCD scales and nuclear 
phenomenology is made by applying 
Georgi and Manohar's Naive Dimensional Analysis (NDA) and 
naturalness \cite{GEORGI84,GEORGI93}.
These principles prescribe how to count powers of
the pion decay constant $f_\pi \approx 94\,$MeV
and
a larger mass scale $\Lambda$
in effective lagrangians or energy functionals.
The mass scale $\Lambda$ is associated with the 
new physics beyond the pions:
the non-Goldstone boson masses or the nucleon mass.
The signature of these low-energy QCD scales in the coefficients of a 
relativistic point-coupling model 
was first pointed out by Friar, Lynn, and Madland \cite{FRIAR96}.
Subsequent analyses have extended and supplemented this idea,
testing it in nonrelativistic mean-field models as well as in different
types of relativistic models.
Estimates of contributions to the energy
functional from individual terms, based on NDA power counting,
are {\em quantitatively\/} consistent with direct, high-quality fits
to bulk nuclear observables \cite{PANIC99,FURNSTAHL00a}.
Naturalness based on NDA scales has proved to be a very robust
concept: nuclei know about these scales!

The successes of QHD mean-field
phenomenology are, at first, rather mysterious
from the EFT perspective alone, since
the Hartree approximation is only the finite-density counterpart of
the Born approximation at zero density.
The density functional theory (DFT) perspective explains 
the successes of mean-field models
and provides a new context for EFT power counting. 

Conventional density functional theory 
is based on energy functionals of the ground-state
density of a many-body system, whose extremization yields a variety of
ground-state properties.
In a covariant generalization of DFT applied to nuclei, 
these become functionals
of the ground-state scalar density $\rhos$ as well as the
baryon current $B_\mu$. 
Relativistic mean-field models are analogs of the Kohn--Sham 
formalism of DFT \cite{KOHN65},
with local
scalar and vector fields $\Phi({\bf x})$ and $W({\bf x})$
appearing in the role of relativistic
Kohn--Sham potentials \cite{SEROT97}.
The mean-field models approximate the exact functional, which includes
all higher-order correlations, using powers and gradients
of auxiliary meson fields or nucleon densities.

The scalar and vector potentials are determined by extremizing the energy
functional,
which gives rise to a Dirac single-particle hamiltonian.
The isoscalar part (for spherical nuclei) is
\begin{equation}
   h_0 = -i\bbox{\nabla\cdot\alpha} + 
        \beta \Bigl(M-\Phi(r)\Bigr)
      +  W(r)  \ , \label{eq:hdirac}
\end{equation}
where $M$ is the nucleon mass
and we define $\Mstar \equiv M - \Phi$.
It is not necessary that $\Phi$ is simply proportional
to a scalar meson field $\phi$.  
In fact, $\Phi$ could be proportional to $\phi$
(as in the original QHD models)
or could be expressed as a sum of scalar and
vector densities (as in relativistic point-couplings models)
or could be a nonlinear function of $\phi$.

Density functional theory can provide a framework for 
the systematic incorporation of correlation effects, 
{\em which are included exactly if the correct functional is identified\/}.
Mean-field models approximate this functional with powers
and gradients of fields
or densities, with the truncation determined by power counting.
The inclusion of vacuum contributions, which was such a difficulty
before, now becomes simple with the realization that all 
necessary counterterms are already present.
The convergence of the EFT/DFT expansion
is reasonable, but is slow enough that there are still too many terms
to calibrate accurately by fitting to nuclear data.
We rely on existing phenomenology as a guide to truncating
the lagrangian most efficiently in light of ill-determined coefficients.
In addition, while it is known that correlation corrections modify the
scalar and vector self-energies by only a small amount (``Hartree 
dominance''), the mean-field energy functional omits possible
nonanalytic terms; a combination of EFT and DFT
may show us how to systematically include them \cite{HU00,DILUTE}.

%\newpage

\subsection{Antidotes, Rebuttals, and Reinterpretations}

In this section, we revisit past criticisms of QHD, many taken 
from earlier {\it Comments\/} \cite{BRODSKY84,NEGELE85,BROWN87}
and others commonly expressed at
conferences or elsewhere in the literature.
We present without attribution a series of paraphrased statements  in
{\bf boldface} that criticize different aspects of QHD and its predictions.
Each statement is followed by a resolution 
based on 
the modern EFT/DFT perspective of QHD.
We find that each criticism is either addressed and answered, revealed
to be incorrect, or rendered moot.

{\bf Nuclei are nonrelativistic systems because corrections to
the kinetic energy are small.}
Relativistic phenomenology for nuclei has often been motivated
by the need for relativistic kinematics when extrapolating to extreme
conditions of density, temperature, or momentum transfer.
Unfortunately, this motivation
obscures the issue of Lorentz covariant vs.\ 
nonrelativistic approaches for nuclei under ordinary conditions.
Relativistic kinematic corrections are indeed small for ordinary
nuclear systems.
The important aspect of relativity in these systems is {\it not\/} 
that a nucleon's momentum is comparable to its rest mass, but that 
maintaining covariance allows scalars to be distinguished from the
time components of four-vectors.
This distinction is easy to see by expanding the self-consistent
nuclear matter energy density $\edens$ in powers of the Fermi momentum
$\kfermi$ (see Ref.~\cite{SEROT97}, p.~554):%
\footnote{Here $\rhoB = 2 \kfermi^3 /3 \pi^2$ 
is the baryon density, and $g_i$ and $m_i$ denote
the scalar and vector couplings and masses, respectively.  
Nonlinear meson interactions are omitted for brevity.}
\beqa
  \edens /\rhoB &=& M + \bigg[ {3\kfermi^2\over 10 M}
            - {3\kfermi^4\over 56 M^3}
            + {\kfermi^6\over 48 M^5} - {15\kfermi^8\over 1408 M^7}
            + {21\kfermi^{10}\over 3328 M^9} + \cdots \bigg]
          \nonumber\\[4pt]
   & & \qquad
      +{\gvsq\over 2\mvsq}\, \rhoB - {\gssq\over 2 \mssq}\, \rhoB
      +{\gssq\over\mssq}\, {\rhoB\over M}\bigg[ {3\kfermi^2\over 10 M}
      -{36\kfermi^4\over 175 M^3} + {16\kfermi^6\over 105 M^5}
      -{64\kfermi^8\over 539 M^7} + \cdots \bigg]
          \nonumber\\[4pt]
  & & \qquad
      + \Big({\gssq\rhoB\over\mssq M}\Big)^{\mkern-2mu 2}
         \bigg[{3\kfermi^2\over 10 M}
          - {351\kfermi^4\over 700 M^3} + \cdots \bigg]
         + \Big({\gssq\rhoB\over\mssq M}\Big)^{\mkern-2mu 3}
            \bigg[{3\kfermi^2\over 10 M}
              - \cdots \bigg] \ + \cdots \ .
        \label{eq:enex}
\eeqa
The corrections to the nonrelativistic kinetic energy (contained in
the first term in brackets) are indeed small at equilibrium density, but
the velocity dependence inherent in the Lorentz scalar interaction 
introduces significant corrections of higher than linear order in $\rhoB$.
The leading correction in each order is {\it repulsive}, so
these corrections are important in establishing the equilibrium point.

In the nuclear medium, a covariant treatment implies distinct
scalar and four-vector nucleon self-energies or optical potentials.
The relevant question is: What are their natural mean values?
QHD phenomenology implies several hundred MeV in the center
of a heavy nucleus.

{\bf The success of nonrelativistic approaches shows that covariant
approaches are wrong or unnecessary.}
Historically, 
the successes of nonrelativistic nuclear phenomenology 
have been cited to cast doubt on the
relevance of large scalar and vector potentials.
But in a nonrelativistic treatment of nuclei, the distinction
between a potential that transforms like a scalar and one that transforms
like the time component of a four-vector is lost.
Because the leading-order contributions of these two types are opposite
in sign,
an underlying large scale characterizing individual covariant potentials
would be {\it hidden\/} in the nonrelativistic central potential.
(The scalar and vector terms add constructively in the nonrelativistic
spin-orbit potential, producing an uncharacteristically large result.)
Furthermore, the EFT expansion implies that
even potentials as large as 300 to 400\,MeV are sufficiently smaller than
the nucleon mass that a nonrelativistic expansion should converge,
if not necessarily optimally.
{\em 
Thus the success of nonrelativistic nuclear phenomenology provides little
direct evidence about covariant potentials.\/}

{\bf The primary focus of nuclear theory is to fit the two-nucleon
data and then to solve the many-body problem.}
The EFT coefficients are fixed by {\em any\/} sufficient
data set; \NN data has no special significance.  
Indeed, a density functional is best determined by
finite-density data.
Furthermore, \NN data by itself {\em cannot\/} be sufficient.
Many-body forces are inevitable \cite{WEINBERG90,DILUTE},
and their size can be estimated
and shown to be non-negligible at ordinary densities (verified 
phenomenologically \cite{FURNSTAHL97,RUSNAK97}).
In a mean-field density functional to be used for medium to heavy nuclei,
at least one three-body and two four-body parameters are 
necessary \cite{FURNSTAHL00a}.

{\bf Why use a field theory?  Potential models are easier.}
Relativistic quantum field theory based
on a local lagrangian density provides a general parametrization
of experimental observables consistent with the essential
physics of the strong interaction: quantum mechanics, special relativity,
unitarity, causality, cluster decomposition, and the intrinsic
symmetries of QCD \cite{WEINBERG95}.
Consequently,
we can describe the physics of low-energy QCD (e.g., ordinary nuclei)
using an effective field theory with hadronic degrees of freedom.
{\it Field theory\/} offers advantages over conventional
{\it potential models\/} by systematically accommodating relativistic
corrections and by allowing the construction of complete,
consistent operators to describe interactions with external probes.

{\bf It is important to find specific observables that distinguish
between relativistic and nonrelativistic theory.}
This pursuit will not be fruitful.
There are field transformations that connect relativistic (covariant)
and nonrelativistic theories, with the ratios of the fields to
the nucleon mass acting as
the parameters controlling truncation (see Refs.~\cite{FOLDY50} and 
\cite{REINHARD89}).
These parameters are small enough that a
nonrelativistic approach should reproduce relativistic results,
although not necessarily at the same level of approximation.
The more appropriate question is:  
What is the most efficient representation?

{\bf Large potentials are an artifact of a relativistic formulation.} 
We argue that the large potentials used in a
covariant description of nuclear phenomenology
are manifestations of the underlying mass scales of low-energy
QCD, which are hidden in nonrelativistic 
treatments \cite{FURNSTAHL00b}.
These QCD mass scales are inescapable if one considers the ${}^1S_0$ NN
phase shift (which becomes repulsive at about 250 MeV laboratory
kinetic energy) together with the singlet scattering length of roughly
$(8\,{\rm MeV})^{-1}$ that signals an almost-bound state near zero 
energy \cite{REID68}.

{\bf It is more efficient to work with a nonrelativistic theory because
large cancellations are built in.}
If there were an approximate
{\it symmetry\/} that enforced the cancellation between
scalar and vector contributions, then it would be desirable to
build the cancellation into any EFT lagrangian or energy functional.
(Chiral symmetry alone does {\it not\/} lead to scalar-vector fine tuning.)
However, if the cancellation is accidental or of 
unknown origin, hiding the underlying scales may be counterproductive.
We argue that nuclei naturally fall into the second category, 
with the relevant scales set not by the nonrelativistic binding
energy and central potential
(tens of MeV), but by the large covariant potentials (hundreds of MeV).
The signals of large underlying scales are patterns in the data
that are simply and efficiently explained by large covariant potentials, but 
which require more complicated explanations in a nonrelativistic 
treatment \cite{FURNSTAHL00b}.
Examples of these are:
\begin{itemize}
\item
The spin-orbit force, which appears automatically
with the observed strength in a covariant formulation, but
is not fully reproduced in even the most sophisticated nonrelativistic
calculations \cite{CARLSON99}.
\item
Medium-energy proton--nucleus spin observables, which are reproduced by
the relativistic impulse approximation with intuitive real
optical potentials \cite{MCNEIL83,SHEPARD83,CLARK83a,CLARK83b}, 
while nonrelativistic
treatments require full-folding and medium effects 
\cite{HYNES85,LUMPE87,RAY90,COKER90},   
and have nonintuitive potentials that change {\em qualitatively\/}
with projectile energy.
\item
The energy dependence of the optical potential for nucleon--nucleus
scattering up to 100\,MeV, which is predicted at the relativistic
mean-field level from the Lorentz structure of the
interaction \cite{SEROT86} (and higher-order corrections are small).
In conventional nonrelativistic treatments, the energy dependence
comes from the nonlocality of exchange
corrections in a Hartree--Fock or Brueckner--Hartree--Fock approximation.
\item
The scalar, isoscalar part of the \NN kernel below 1\,GeV, which can be
studied in an essentially model-independent way.
Chiral symmetry, unitarity, and the natural strength of the
$\pi\pi$ interaction imply an integrated strength that results in
a large scalar single-particle potential \cite{LIN89}.
We are not aware of a loophole here.
\item
The equilibrium of nuclear matter, which is not an ordinary, 
nonrelativistic Fermi liquid, since 
it is too dilute and too weakly bound.
These characteristics arise in the mean-field
energy/particle from an empirically small
coefficient of the $\kfermi^3$ term in a density expansion\ 
[see Eq.~(\ref{eq:enex})].
The cancellation between scalar and vector contributions 
to the nuclear matter binding energy account for this fine tuning. 
\end{itemize}

{\bf There is no experimental evidence of large scalar and 
vector fields.}
There can be no {\it direct\/} experimental verification
(or refutation) of any nuclear potentials.
The evidence that a natural representation contains large fields, which is 
achieved only with a covariant formulation,
comes from both empirical and theoretical analyses of \NN scattering and
nuclear properties \cite{MACHLEIDT89,FURNSTAHL00a,FURNSTAHL00b}.  
As noted above, this manifestation of QCD scales translates
in many instances into simpler, more efficient, more compelling
explanations of nuclear phenomena than in nonrelativistic formulations.

%%%%%%%%%%%%%%%%%%%%%%%%%%%%%%%%%%%%%%%%%%%%%%%%%%%%%%%%%%
\begin{figure}[t]
\begin{center}
%\hbox{
\epsfxsize=3.15in
\epsffile{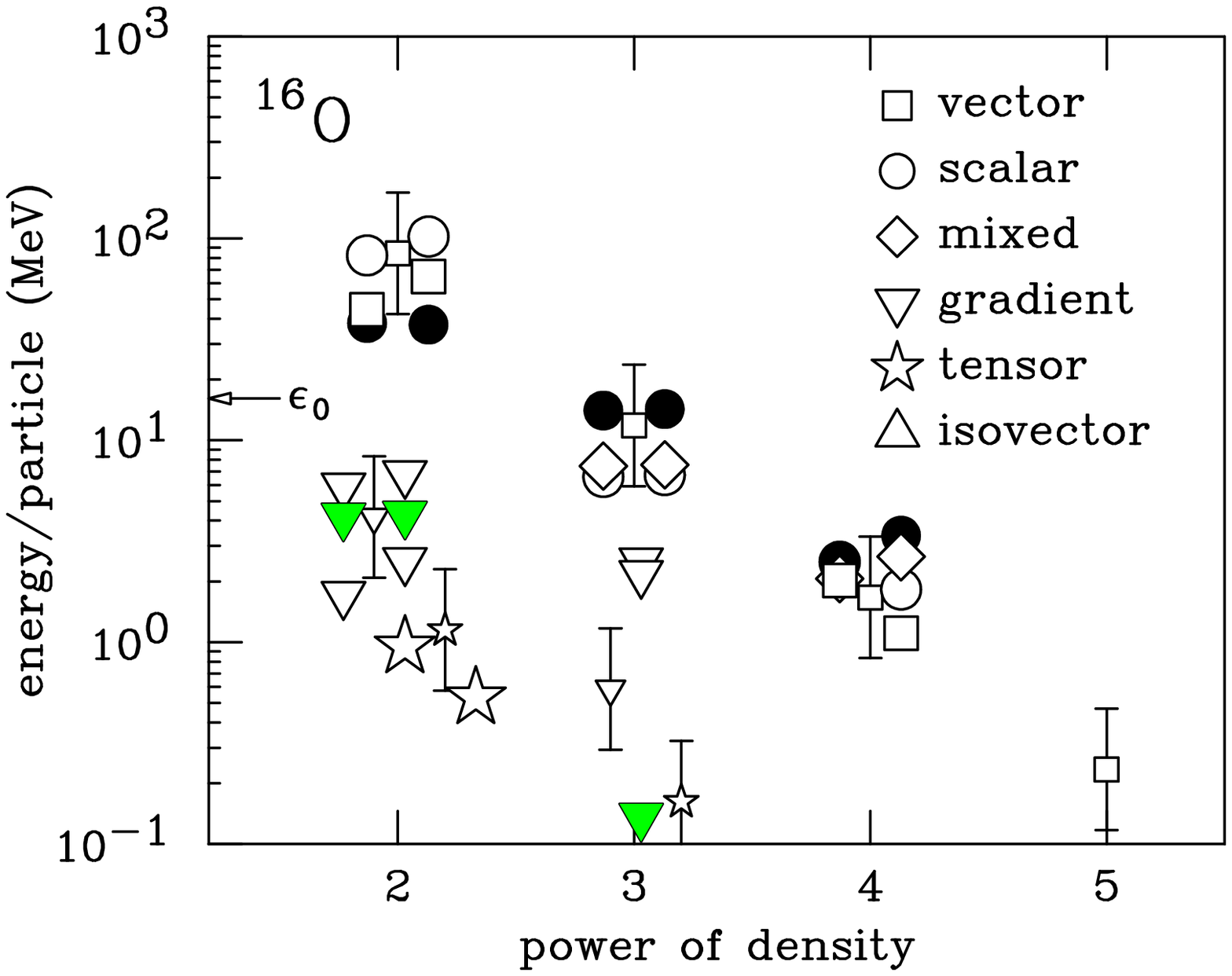}
%\hspace*{.1in}
\hfill
\epsfxsize=3.15in
\epsffile{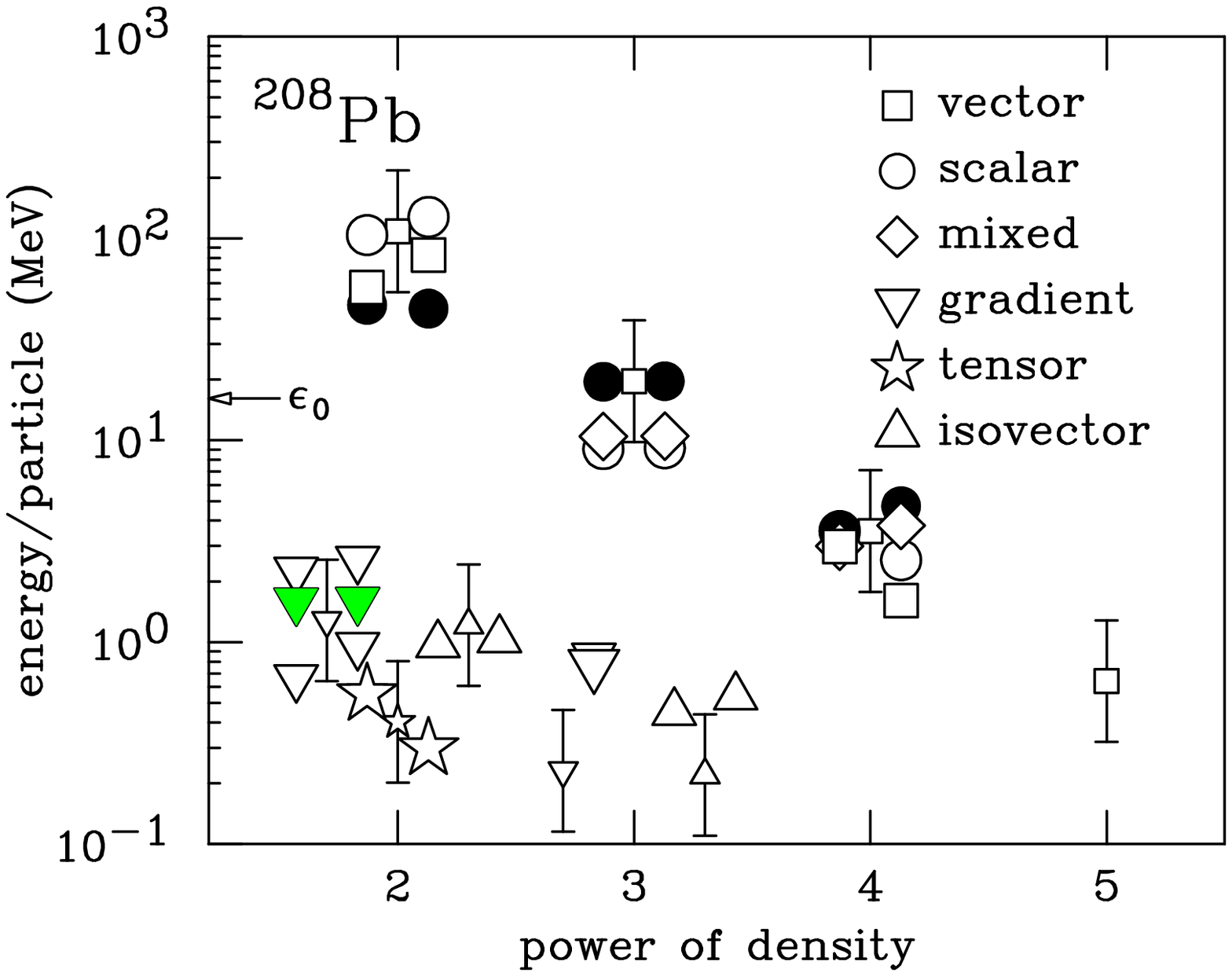}
\end{center}
%\vspace*{-.2in}
\caption{Contributions to the energy/particle in ${}^{16}$O
and ${}^{208}$Pb
 for two covariant, mean-field,
 point-coupling models \protect\cite{FURNSTAHL00a}.
 Absolute values are shown.
 The filled symbols are net values. 
The small symbols indicate estimates based on NDA, with the error bars
corresponding to natural coefficients from 1/2 to 2.
The equilibrium binding energy of nuclear matter is 
$\epsilon_{\scriptscriptstyle 0}$.}
\label{fig:o16energy1}
\end{figure}
%%%%%%%%%%%%%%%%%%%%%%%%%%%%%%%%%%%%%%%%%%%%%%%%%%%%%%%%%%%%%%

Empirical support from nuclear properties comes from the study of
covariant density functionals fit to nuclei \cite{FURNSTAHL97}.
A good fit to nuclear properties requires the local scalar and vector 
potentials to be roughly 300\,MeV, and the hierarchy of energy
contributions follow NDA predictions \cite{FURNSTAHL00a} (see Fig.~1).
A more subtle argument is that the spread of 15\,MeV or more
among ``realistic'' nonrelativistic predictions of the nuclear
matter equilibrium binding energy 
(the ``Coester line'') \cite{MACHLEIDT89}
would be difficult to understand as calibration errors
(``off-shell effects''),
if the underlying scale of the two-body interaction were only 50\,MeV.
In contrast, large covariant two-body potentials in a relativistic
formulation imply sizable three-body contributions in the
corresponding nonrelativistic calculation that are consistent with
this spread.
Finally, independent empirical support comes from fits of
a covariant kernel for the \NN interaction, which is used to calculate the
\NN scattering matrix.  
Every accurate fit has led to large, isoscalar,
scalar and vector contributions of comparable magnitude, but of opposite
sign, which translate in the medium into single-particle potentials
of several hundred MeV at equilibrium density \cite{MACHLEIDT89}. 

The pieces of evidence supporting a representation with large nucleon 
scalar and vector potentials, while not definitive when considered
individually, collectively comprise a compelling positive 
argument \cite{FURNSTAHL00b}.

{\bf We know that mean-field theory cannot be a correct description
of nuclei because important long- and short-range correlations are omitted.}
Mean-field models are approximate implementations of Kohn--Sham
density functional theory \cite{KOHN65}, which means that
correlation effects are included in simple Hartree calculations.
Moreover, the ``Hartree dominance'' of the single-particle potentials has
been demonstrated, implying that short-range correlation corrections are no
more than tens of MeV \cite{HOROWITZ84,HOROWITZ87,SEROT86}.
The bulk properties of interest in mean-field phenomenology are primarily
isoscalar observables that involve low resolution \cite{FURNSTAHL00a}, so
long-range pionic correlations are of minor importance; for other
observables, EFT provides a systematic framework for explicitly including
pionic contributions.

{\bf The relativistic mean-field approximation may be valid at high densities, 
but nuclei are low-density systems.}
The modern view is that the successes of relativistic mean-field theory
do not depend on the justification of the mean-field approximation
at high density, but on the flexibility of the mean-field 
density functional near equilibrium density.
The combination of EFT and DFT concepts and methods applied to mean-field
models of nuclei reveals that:
\begin{itemize}
  \item NDA provides an organizational principle for the EFT.
  Power counting and the limited number of bulk nuclear observables
  explain the success of conventional mean-field models, which contain
  fewer parameters than the most general EFT models.
  \item Vacuum effects, chiral symmetry, and nucleon substructure
  {\it are all included\/} in general QHD models.  
  \item Ground-state nuclear properties provide information at
  low resolution.
  Models with different degrees of freedom (e.g., four- vs.\ 
  two-component
  nucleons or point-coupling vs.\ meson models) 
  are simply different organizations of the EFT.  All are
  consistent with NDA. 
\end{itemize}

{\bf Relativistic many-body calculations have unquantifiable errors.}
The EFT framework based on NDA and naturalness
provides an organizational scheme for truncating a lagrangian
or energy functional and for making well-defined error 
estimates (see Fig.~1).
Vacuum corrections, which disrupted early attempts at QHD expansion schemes,
are innocuous in the EFT approach.
(They are automatically absorbed into the coefficients.)
Moreover, there are (in principle) no off-shell ambiguities.

{\bf QHD calculations apply perturbation theory, which 
is not sensible with large coupling
constants.} 
In fact, QHD does have a sensible expansion, which is 
{\it not\/} in powers of the couplings.
We work instead with density functional theory, with NDA power
counting identifying reasonable expansion parameters.  
It is true that the short-distance (ultraviolet) behavior may be
incorrect, but the EFT can correct the behavior systematically
for low-energy observables using a small number of parameters
(verified phenomenologically \cite{FURNSTAHL00a}).

{\bf Calculations of magnetic moments in relativistic models 
are inconsistent with the data due to enhancements from a
small effective nucleon mass.}
Naively, the baryon current of a nucleus with a single valence nucleon 
with momentum $p$ 
outside a closed shell is $p/\Mstar$, compared to the Schmidt current
$p/M$.
However, if the calculation is forced to respect Lorentz covariance
and the first law of 
thermodynamics, the nuclear current is constrained to be $p/\mu$,
where $\mu \approx M$ is the chemical potential 
\cite{FURNSTAHL87}.
Thus there is no enhancement in a consistent relativistic framework.

{\bf Relativistic theories have ghosts.}
No they don't.
Ghosts arise from improper treatment of the ultraviolet behavior in
renormalizable theories; in EFT, this short-distance behavior
is included systematically by fitting a small number of parameters
to nuclear observables.
Moreover, the long-distance instability known as 
``Brown--Ravenhall disease'' \cite{BROWN51}
does not arise in EFT, because the EFT framework is not
quantum mechanics with a fixed number of particles.

{\bf The successes of the relativistic impulse approximation
are meaningless because correlation corrections are large.}
On the contrary, relativistic correlation corrections to the optical
potential are small, in contrast to the nonrelativistic framework 
(``Hartree dominance'').
Thus the success of RIA calculations is expected.
We emphasize that covariant and nonrelativistic expansions
can have very different rates of convergence.
The covariant representation appears superior in most instances
\cite{HYNES85,COKER90}.

{\bf G-parity implies that a successful description of \NN scattering
leads to unphysical predictions in the $\overline{\bf N}$N sector.}
Explicit calculations show that absorptive processes dominate
the \NNbar optical potential \cite{CLARK84b}.
Thus the consequences of $G$-parity are not directly observable, since
the transformed \NN scattering amplitude is known only at unphysical
kinematics for the \NNbar system.
Moreover, in the EFT framework, there are no on-shell antinucleons;
there are only valence nucleons.
A simple extension of QHD to the \NNbar sector pushes the EFT expansion
beyond its breakdown scale $\Lambda$, so this extrapolation is suspect. 

{\bf QHD does not have pions and chiral symmetry.}
Modern QHD effective lagrangians include pions in a nonlinear 
realization of chiral symmetry \cite{FURNSTAHL97}.
Confusion about the apparent absence of pions arises because
{\it explicit\/} pionic contributions do not contribute to mean-field
energy functionals, although correlated pionic contributions are 
{\it implicitly\/} contained in the effective scalar field.
Long-range pionic contributions can be included systematically,
but do not qualitatively change mean-field
phenomenology \cite{HOROWITZ83,SEROT86,HU00}.

{\bf The factorization of nuclear amplitudes into a product of
on-shell, single-nucleon form factors and many-body amplitudes is incorrect.}
The modern chiral EFT lagrangians of QHD do not require such a factorization.
The single-nucleon structure is included explicitly in the lagrangian
through a derivative expansion \cite{FURNSTAHL97}.
This produces results that are very similar to the standard ``folding''
procedure \cite{HOROWITZ81}.

{\bf The anomalous moment of the nucleon is clearly a property
of its internal quantum structure; by itself, this precludes the
representation of the nucleon as a local field.}
This is directly refuted by the EFT lagrangian \cite{FURNSTAHL97}, 
which not only accommodates an anomalous moment, it requires it!

{\bf Local meson fields and ``point'' nucleons provide no possibility
for quark substructure.}
This is simply incorrect. 
At energy and momentum scales small compared to the underlying QCD scale
$\Lambda$, details of the quark substructure are not resolved.
It follows that the substructure can be incorporated 
through a systematic expansion of nonlinear and gradient interactions in the
effective lagrangian, with the dynamics
encoded in the local hadronic couplings.
This is the essence of the EFT approach.
A clear example of this expansion is the single-nucleon structure
included in modern QHD chiral lagrangians \cite{FURNSTAHL97}.

{\bf Virtual nucleon--antinucleon Z graphs, which are essential
to relativistic phenomenology, should be suppressed because
the nucleon has substructure.}
A local Dirac field for the nucleon does not imply a {\em physical\/}
point nucleon \cite{SEROT86,FURNSTAHL97}.
Moreover, the virtual \NNbar pair is {\it far off shell}, and
the off-shell intermediate states in a particular representation
cannot be interpreted in terms of on-shell physics.
The EFT framework ensures that any incorrect short-distance dynamics
can be corrected systematically with counterterms.
Recent formulations of covariant chiral perturbation theory also verify
that implicit Z graphs are not a problem, and 
that consistent power counting is possible \cite{ELLIS98,BECHER99}.

{\bf The QHD treatment of the vacuum neglects nucleon substructure,
violates $N_c$ counting rules, and relies on unphysical
$\overline{\bf N}$N contributions (Z graphs).}
The modern QHD treatment of vacuum dynamics has changed this
discussion completely. 
Vacuum contributions in the EFT framework are not calculated
explicitly but are implicitly and systematically
contained in a small number of fitted parameters.
Any physical consequences of hadronic substructure, for example,
are automatically included.
Furthermore, these implicit contributions are consistent with $N_c$ counting
using NDA \cite{RHA97}.

{\bf Hidden QCD color is relevant for low-energy nuclear physics and is
not contained in QHD \cite{BRODSKY84}.}
The EFT perspective says this objection 
{\it must be irrelevant}, because all
observable amplitudes are color singlets (since color is confined).  
Regardless of the underlying QCD picture, the EFT must be valid at
sufficiently low energies, without explicitly invoking
colored degrees of freedom.

{\bf Quantum chromodynamics of quarks and gluons is the 
fundamental theory of the strong interaction,
so we should describe nuclei in terms of quarks.}
For most ordinary nuclear phenomena, a description based on hadronic
degrees of freedom is most appropriate: Hadrons are the particles 
actually observed in experiments and thus are more efficient. 
Hadronic calculations can be calibrated using empirical nuclear properties
and scattering observables.
Hadronic models have historically provided accurate descriptions of \NN
scattering and the bulk and single-particle properties of nuclei.
It is better to {\it match\/} the effective hadronic theory to QCD
to determine its coefficients and then use the EFT to calculate
nuclear structure and reactions.

%\newpage

\section{Discussion}

\subsection{Summary}

The hadronic theory of QHD is truly a manifestation of QCD in the
strong-coupling regime.
It is currently impossible to construct this theory directly from
QCD, but the effective field theory perspective shows that we can make
progress regardless, without recourse to {\it ad hoc\/} models.  

The history of QHD from 1974 shows an evolution driven by the successes and
difficulties of the original approach.
One of the cornerstones of this approach was the necessity for a consistent,
microscopic treatment of nuclear systems using hadrons \cite{WALECKA74}.
Although the mean-field theory was phenomenologically successful, the
ultimate goal was to improve upon this approximation to incorporate both
many-body and short-distance (quantum vacuum) effects.
Unfortunately, this goal was sometimes overlooked, because numerous
``improvements'' degraded the quality of the mean-field-theory results;
this often led to the imposition of arbitrary constraints or restrictions
on QHD calculations.
Fortunately, however, perseverance within the original framework ultimately
led to the conclusion that the constraint of renormalizability is too
restrictive, and alternatives to this requirement were sought.

The result is the modern viewpoint of QHD based on effective field theory
and density functional theory (the ``revolution''), which solves 
the most serious problems while preserving intact the successful
predictions for bulk and single-particle nuclear observables.
The EFT framework identifies the systematic ``organizing principle''
behind the successful QHD calculations:  energy scales arising from the
underlying QCD define the dimensional analysis for terms in the effective
lagrangian.
Naturalness and the size of the nuclear mean fields
allow for a practical expansion and truncation,
and also clarify the scope and limitations of QHD.
Density functional theory and the Kohn--Sham formalism then explain why
the truncated mean-field energy functional can be flexible enough to
yield accurate results for certain nuclear observables.

Analyses based on QHD, as defined here, provide a
correct description of baryonic systems at sufficiently 
large distances and low energies.
But we argue further that
a {\it covariant\/} formulation of the dynamics manifests
the {\it true energy scales\/} of QCD in nuclei and provides an 
efficient and comprehensive explanation of observed bulk
and single-particle systematics.

\subsection{Outlook}

The development of QHD is far from over, and there are many issues to
be addressed.
Whereas the majority of existing QHD calculations focus on isoscalar physics,
the incorporation of pions using a nonlinear realization of chiral symmetry
and the inclusion of the Delta baryon as a collective $\pi$N degree of
freedom should produce accurate results in the isovector sector; this
merits further study.
Calculations of excited states using a consistent, conserving approach to
the random-phase approximation \cite{JORGEP00},
coupled with new, more accurate data on nuclear breathing 
modes \cite{GMR99}, could
lead to a more precise determination of QHD parameters \cite{FURNSTAHL00a}.
The situation for nuclear currents and magnetic nuclear form factors
at low momentum transfer is still an open problem that must be
re-examined in the context of modern EFT.
It is also important to pursue the connection between QHD results for
many-body systems and the recent calculations of few-nucleon systems
within the EFT framework, including covariant chiral perturbation theory.
A major challenge is to develop and apply systematic and consistent 
``power counting'' schemes that lead to more general
conserving approximations and to study renormalization-group methods
that could determine the analytic structure of the ground-state
energy functional \cite{DILUTE}.
In addition, we must learn how to extrapolate beyond the breakdown scale
of the EFT description.

Ultimately we must answer the question:
What is the best way to connect nuclear phenomenology to QCD?  
Quark models have been advocated, but they continue to lack a
systematic framework and a direct connection to QCD.
The EFT perspective makes explicit quark degrees of freedom nonessential. 
Nevertheless, finding an efficient, tractable, nonperturbative way to
match the QCD lagrangian to the long-range, strong-coupling, effective
field theory of QHD is a major goal for the future.

\bigskip
\acknowledgments

We thank B.\ C.\ Clark, H.-W.\ Hammer,
D.\ B.\ Lichtenberg, and J.\ D.\ Walecka
for useful discussions and comments.
This work was supported in part by the National Science Foundation under
Grant No.\ PHY-9800964 and by the U.S. Department of Energy under
Contract No. DE-FG02-87ER40365.


\begin{references}
%
\bibitem{SCHIFF51}L. I. Schiff, Phys.\ Rev.\ {\bf 84} (1951) 1, 10.
%
\bibitem{WALECKA74}J. D. Walecka, Ann.\ Phys.\ (N.Y.) {\bf 83} (1974) 491.
%
\bibitem{SEROT86}B.~D.\ Serot and J.~D.\ Walecka, Adv.\ Nucl.\
    Phys.\ {\bf 16} (1986) 1.
%
\bibitem{SEROT91} B.~D.\ Serot and J.~D.\ Walecka, Recent Progress
    in Many-Body Theories, vol. 3, T.~L.\ Ainsworth, 
    C.~E.\ Campbell, B.~E.\ Clements, and E.\ Krotscheck, eds.
    (Plenum, New York, 1992), p.~49.
%
\bibitem{MUELLER96} H. M{\"u}ller and B. D. Serot, Nucl.\ Phys.\ 
    {\bf A606} (1996) 508.
%
\bibitem{HOROWITZ81} C. J. Horowitz and B. D. Serot, Nucl.\ Phys.\ 
    {\bf A368} (1981) 503.
%
\bibitem{REINHARD86} P.-G.\ Reinhard, M.\ Rufa, J.\ Maruhn, W.\ Greiner,
        and J.\ Friedrich, Z.\ Phys. {\bf A323} (1986) 13.
%
\bibitem{RUFA88} M.\ Rufa, P.-G.\ Reinhard, J.~A.\ Maruhn, W.\ Greiner,
        and M.~R.\ Strayer, Phys.\ Rev. C {\bf 38} (1988) 390.
%
\bibitem{REINHARD89} P.-G.\ Reinhard, Rep.\ Prog.\ Phys.\ {\bf 52}
        (1989) 439.
%
\bibitem{GAMBHIR90} Y.~K.\ Gambhir, P.\ Ring, and A.\ Thimet, Ann.\ Phys.\ 
        (N.Y.) {\bf 198} (1990) 132.
%
\bibitem{SEROT92} B. D. Serot, Rep.\ Prog.\ Phys.\ {\bf 55} (1992) 
      1855.
%
\bibitem{SEROT97}B.~D.\ Serot and J.~D.\ Walecka, Int.\ J.\ Mod.\
    Phys.\ E {\bf 6} (1997) 515.
%
\bibitem{MCNEIL83} J. A. McNeil, J. R. Shepard, and S. J. Wallace,
    Phys.\ Rev.\ Lett.\ {\bf 50} (1983) 1439.
%
\bibitem{SHEPARD83} J. R. Shepard, J. A. McNeil, and S. J. Wallace,
    Phys.\ Rev.\ Lett.\ {\bf 50} (1983) 1443.
%
\bibitem{CLARK83a} B. C. Clark, S. Hama, R. L. Mercer, L. Ray, and 
    B. D. Serot, Phys.\ Rev.\ Lett.\ {\bf 50} (1983) 1644.
%
\bibitem{CLARK83b} B. C. Clark, S. Hama, R. L. Mercer, L. Ray,
    G. W. Hoffmann, and B. D. Serot, Phys.\ Rev.\ C {\bf 28} (1983) 1421.
%
\bibitem{CLARK84a} B. C. Clark, S. Hama, E. Sugarbaker, M. A. Franey,
    R. L. Mercer, L. Ray, G. W. Hoffmann, and B. D. Serot,
    Phys.\ Rev.\ C {\bf 30} (1984) 314.
%
\bibitem{DAWSON90} J.~F.\ Dawson and R. J. Furnstahl, Phys.\ Rev.\ C
        {\bf 42} (1990) 2009.
%
\bibitem{BROCKMANN84} R. Brockmann and R. Machleidt, Phys.\ Lett.\ 
       {\bf 149B} (1984) 283; Phys.\ Rev.\ C {\bf 42} (1990) 1965.
%
\bibitem{MACHLEIDT89} R. Machleidt, Adv.\ Nucl.\ Phys.\ {\bf 19} (1989) 189.
%
\bibitem{DEJONG91} F.~de Jong and R.\ Malfliet, Phys.\ Rev.\ C {\bf 44}
        (1991) 998.
%
\bibitem{HOROWITZ84} C. J. Horowitz and B. D. Serot, Phys.\ Lett.\ 
        {\bf 137B} (1984) 287.
%
\bibitem{HOROWITZ87} C. J. Horowitz and B. D. Serot, Nucl.\ Phys.\ 
        {\bf A464} (1987) 613; {\bf A473} (1987) 760 (E).
%
\bibitem{TWOLOOP89} R. J. Furnstahl, R. J. Perry, and B. D. Serot,
        Phys.\ Rev.\ C {\bf 40} (1989) 321.
%
\bibitem{MILANA91} J. Milana, Phys.\ Rev.\ C {\bf 44} (1991) 527.
%
\bibitem{ALLENDES92} M. P. Allendes and B. D. Serot, Phys.\ Rev.\ C
        {\bf 45} (1992) 2975.
%
\bibitem{SCHWINGER57} J. Schwinger, Ann.\ Phys.\ (N.Y.) {\bf 2}
        (1957) 407.
%
\bibitem{SIGMA60} M. Gell-Mann and M. L{\'e}vy, Nuovo Cim. {\bf 16}
        (1960) 705.
%
\bibitem{FURNSTAHL93a} R. J. Furnstahl and B. D. Serot, Phys.\ Rev.\ C
        {\bf 47} (1993) 2338.
%
\bibitem{FURNSTAHL93b} R. J. Furnstahl and B. D. Serot,
        Phys.\ Lett.\ B {\bf 316} (1993) 12.
%
\bibitem{FURNSTAHL96} R. J. Furnstahl, B. D. Serot, and H.-B. Tang, 
          Nucl.\ Phys.\ {\bf A598} (1996) 539.
%
\bibitem{WEINBERG67} S. Weinberg, Phys.\ Rev.\ Lett.\ {\bf 18} (1967)
         188; Phys.\ Rev.\ {\bf 166} (1968) 1568.
%
\bibitem{JENKINS91} E. Jenkins and A. V. Manohar, Phys.\ Lett.\ B
         {\bf 255} (1991) 558.
%
\bibitem{ELLIS98} P. J. Ellis and H.-B. Tang, Phys.\ Rev.\ C {\bf 57}
    (1998) 3356.
%
\bibitem{BECHER99} T. Becher and H. Leutwyler, Eur.\ Phys.\ J.\ C {\bf 9}
    (1999) 643.
%
\bibitem{FURNSTAHL97} R. J. Furnstahl, B. D. Serot, and H.-B. Tang, 
        Nucl.\ Phys.\ {\bf A615} (1997) 441.
%
\bibitem{GEORGI84} H. Georgi and A. Manohar, Nucl.\ Phys.\ 
        {\bf B234} (1984) 189.
%
\bibitem{GEORGI93} H. Georgi, Phys.\ Lett.\ B {\bf 298} (1993) 187.
%
\bibitem{FRIAR96} J. L. Friar, D. G. Madland, and B. W. Lynn,
    Phys.\ Rev.\ C {\bf 53} (1996) 3085.
%
\bibitem{PANIC99}R.~J.\ Furnstahl and B.~D.\ Serot,
   Proc.\ $15^{\rm th}$ Particles and Nuclear International Conference
   (PANIC99), Uppsala, Sweden, Nucl.\ Phys.\ {\bf A663} and {\bf A664}
   (2000) 513c.
%      e-print nucl-th/9907073.
%
% parameters paper
\bibitem{FURNSTAHL00a}R.~J.\ Furnstahl and B.~D.\ Serot,                                           
  Nucl.\ Phys.\ {\bf A671} (2000) 447.
%
\bibitem{KOHN65} W.\ Kohn and L.~J.\ Sham, Phys.\ Rev.\ {\bf A140} (1965)
       1133.
%
\bibitem{HU00} Y.\ Hu, Ph.D. thesis, Indiana University (2000).
%
\bibitem{DILUTE} H.-W. Hammer and R. J. Furnstahl, 
       {\tt [arXiv:nucl-th/0004043 v2]}, Nucl.\ Phys.\ A, in press.
%
\bibitem{BRODSKY84}S.~J.\ Brodsky, Comm.\ Nucl.\ Part.\ Phys.\ 
      {\bf 12} (1984) 213.
%
\bibitem{NEGELE85}J.~W.\ Negele, Comm.\ Nucl.\ Part.\ Phys.\ 
           {\bf 14} (1985) 303.
%
\bibitem{BROWN87}G.~E.\ Brown, W.~Weise, G.~Baym, and J.~Speth,
     Comm.\ Nucl.\ Part.\ Phys.\ {\bf 17} (1987) 39.
%
\bibitem{WEINBERG90} S. Weinberg, Phys.\ Lett.\ B {\bf 251} (1990) 288;
     Nucl.\ Phys.\ B {\bf 363} (1991) 363.
%
\bibitem{RUSNAK97} J. J. Rusnak and R. J. Furnstahl, 
     Nucl.\ Phys.\ {\bf A627} (1997) 495.
%
\bibitem{WEINBERG95} S. Weinberg, The Quantum Theory of Fields, vol. I
      (Cambridge Univ.\ Press, Cambridge, UK, 1995).
%
\bibitem{FOLDY50} L. L. Foldy and S. A. Wouthuysen, Phys.\ Rev.\ {\bf 78}
      (1950) 29.
%                  
% large potentials paper                                                             
\bibitem{FURNSTAHL00b}R.~J.\ Furnstahl and B.~D.\ Serot,                                           
%    {\tt [arXiv:nucl-th/9912048]}, 
     Nucl.\ Phys.\ {\bf A673} (2000) 298.
%
\bibitem{REID68} R. V. Reid, Jr., Ann.\ Phys.\ (N.Y.) {\bf 50}
    (1968) 411.
%
\bibitem{CARLSON99} J. Carlson, Few-Body Systems Suppl. {\bf 10} (1999) 1.
%
\bibitem{HYNES85} M. V. Hynes, A. Picklesimer, P. C. Tandy, and R. M. Thaler,
    Phys.\ Rev.\ C {\bf 31} (1985) 1438.
%
\bibitem{LUMPE87} J. D. Lumpe and L. Ray, Phys.\ Rev.\ C {\bf 35} 
    (1987) 1040.
%
\bibitem{RAY90} L. Ray, Phys.\ Rev.\ C {\bf 41} (1990) 2816.                                                                               
%
\bibitem{COKER90} W. R. Coker and L. Ray, Phys.\ Rev.\ C {\bf 42} (1990) 659.
%
\bibitem{LIN89} W. Lin and B. D. Serot, Phys.\ Lett.\ B {\bf 233} (1989)
    23; Nucl.\ Phys.\ {\bf A512} (1990) 637.
%
\bibitem{FURNSTAHL87}R.~J.\ Furnstahl and B.~D.\ Serot,
    Nucl.\ Phys.\ {\bf A468} (1987) 539;
    Phys.\ Rev.\ C {\bf 43} (1991) 105.
%
\bibitem{BROWN51} G. E. Brown and D. G. Ravenhall, Proc.\ Roy.\ Soc.\
    London, {\bf A208} (1951) 552.
%
\bibitem{CLARK84b} B. C. Clark, S. Hama, J. A. McNeil, R. L. Mercer, L. Ray,
    B. D. Serot, D. A. Sparrow, and K. Stricker-Bauer,
    Phys.\ Rev.\ Lett.\ {\bf 53} (1984) 1423.
%
\bibitem{HOROWITZ83} C. J. Horowitz and B. D. Serot, Nucl.\ Phys.\ 
    {\bf A399} (1983) 529.
%
\bibitem{RHA97} R. J. Furnstahl, B. D. Serot, and H.-B. Tang, 
    Nucl.\ Phys.\ {\bf A618} (1997) 446.
%
\bibitem{JORGEP00} J. Piekarewicz, {\tt [arXiv:nucl-th/0003029]}.
%
\bibitem{GMR99} D. H. Youngblood, H. L. Clark, and Y.-W. Lui, 
    Phys.\ Rev.\ Lett.\ {\bf 82} (1999) 691.
%
\end{references}
\end{document}